\documentclass[10pt]{JHEP3}
\pdfoutput=1
\usepackage{amsmath}
\usepackage{verbatim}
\usepackage{graphicx}

\newcommand{\e}{\epsilon}

\renewcommand{\L}{{\mathcal{L}}}
\newcommand{\bL}{\bar{{\mathcal{L}}}}
\newcommand{\z}{{\bar z}}

\newcommand{\be}[1]{ \begin{equation}\label{#1} }
\newcommand{\ee}{\end{equation}}
\newcommand{\ben}[1]{\begin{eqnarray}\label{#1} }
\newcommand{\een}{\end{eqnarray}}

\newcommand{\p}{\partial}

\newcommand{\refb}[1]{(\ref{#1})}
\newcommand{\w}{\omega}
\newcommand{\bw}{\bar{\omega}}

\renewcommand{\>}{\rangle}

\renewcommand{\z}{{\bar z}}
\newcommand{\h}{{\bar h}}

\DeclareMathOperator{\extdm}{d}
\newcommand{\extd}{\extdm \!}

\title{3D Flat Holography: Entropy and Logarithmic Corrections.}

\author{Arjun Bagchi 
\\
$\;$ $\,$ Indian Institute of Science Education and Research \\
$\;$ $\,$ Dr Homi Bhabha Road, Pashan. Pune 411008. INDIA \\

$\;$\email{a.bagchi@iiserpune.ac.in}
}

\author{Rudranil Basu 
\\
$\;$ $\,$ Centre for High Energy Physics, Indian Institute of Science \\
$\;$ $\,$ C.V. Raman Avenue, Bangalore 560012, INDIA. \\

$\;$\email{rudranil@cts.iisc.ernet.in}}
\abstract{We compute the leading corrections to the Bekenstein-Hawking entropy of the Flat Space Cosmological (FSC) solutions in 3D flat spacetimes, which are the flat analogues of the BTZ black holes in AdS$_3$. The analysis is done by a computation of density of states in the dual 2D Galilean Conformal Field Theory and the answer obtained by this matches with the limiting value of the expected result for the BTZ inner horizon entropy as well as what is expected for a generic thermodynamic system. Along the way, we also develop other aspects of holography of 3D flat spacetimes.}

\preprint{}

\begin{document}

\baselineskip 3.5ex

\section{Introduction}
\noindent
{
The Holographic Principle \cite{'tHooft:1993gx, Susskind:1994vu} provides us with a unique route of understanding quantum gravity by looking at a theory without gravity. Its best understood avatar, viz. the AdS/CFT correspondence \cite{Maldacena:1997re}, has been one of the most successful tools of modern theoretical physics. One of the celebrated successes of AdS/CFT is the explanation of the entropy of black holes in terms of the dual field theory \cite{Strominger:1996sh}. 

Gravity in three dimensions has long been an attractive testing ground for attempts at a quantum theory of gravity. The lack of propagating degrees of freedom makes life simple in three dimensions and gravity can be formulated equivalently as a Chern-Simons theory \cite{Witten:1988hc, ach-town}. One of the main pre-cursers of AdS/CFT was the analysis of asymptotic symmetries of AdS$_3$ by Brown and Henneaux \cite{bh} who found that the Asymptotic Symmetry Algebra (ASA) enlarges from the isometry algebra of $SO(2,2)$ to two copies of the Virasoro algebra. The power of these symmetries of a 2d Conformal Field Theory allows one to compute many quantities without having knowledge of the specific details of the dual theory like a specific Lagrangian. 

One of the great surprises of gravity in AdS$_3$ was the discovery of the BTZ black hole solutions  \cite{Banados:1992wn, Banados:1992gq} despite the absence of any propagating degrees of freedom. In this decade of holographic studies, the BTZ black holes have been extensively used to understand many aspects of AdS/CFT. One of the very early tests was the matching of the entropy of the BTZ black hole to a counting of states in 2d CFT \cite{Strominger:1997eq}. The power of the infinite symmetries was brought to use here and by resorting to Cardy's famous analysis \cite{Cardy:1986ie} which invokes general properties of 2d CFT like modular invariance, the bulk Bekenstein-Hawking was matched to the boundary analysis. 

The expectation from the gravitational side is that the leading correction to the Bekenstein-Hawking entropy would go like the logarithm of the area of the event horizon and the next order would be proportional to the inverse of the area and so on. For the BTZ black hole, this has been shown from a CFT calculation in \cite{Carlip:2000nv}. The leading quantum correction to the black hole entropy can be calculated by extending Cardy's trick to the next order. 

The holographic principle is a general statement about the equivalence of the gravitational theory to a lower dimensional field theory without gravity and thus should hold beyond the known and much-explored example of AdS/CFT. Attempts at understanding a de Sitter analogue of this correspondence has been made by analytic continuations from AdS \cite{Strominger:2001pn}. For more recent developments, the reader is referred to the review \cite{Anninos:2012qw} and the references therein.  Flat spacetimes have been far less explored from a holographic view point. Interestingly, asymptotic symmetry structures were known in 4d flat spacetimes much before the seminal analysis of Brown and Henneaux and it was known that an infinite dimensional algebra, called the Bondi-Metzner-Sachs (BMS) algebra, appears at the null-boundary of flat 4d spacetimes \cite{Bondi:1962px, Sachs:1962zza}. In 3d as well, infinite dimensional ASA's were discovered \cite{Barnich:2006av} and this is called the BMS$_3$ algebra. Surprisingly, this algebra had also been studied as the symmetries of a 2d non-relativistic conformal field theory or a Galilean Conformal Field Theory \cite{Bagchi:2009my, Bagchi:2009pe} and this connection was dubbed the BMS/Galilean Conformal Algebra correspondence \cite{Bagchi:2010eg}. Given that flat spacetimes arise in the large radius limit of AdS, it is natural to try and formulate flat holography as a limit of AdS/CFT. It was shown that the flat space symmetry structure could be derived as a limit of the AdS$_3$ symmetry structure \cite{{Bagchi:2012cy, Barnich:2012aw}}. There have been several follow-up works to this \cite{Bagchi:2012yk} -- \cite{Fareghbal:2013ifa} building on aspects of holography in 3D flat spacetimes.  

A natural question that arises in this context is what happens if one looks to take the flat limit on a BTZ black hole \cite{Bagchi:2012xr}. It is well known that 3d flat spacetimes don't allow any black hole solutions, so what happens is quite exotic. If one starts off with a non-extremal BTZ black hole which has an outer and an inner horizon, the outer horizon moves out to infinity in this limit and one is left with the inside of the original black hole. Here temporal and radial directions flip and what used to be a black hole solution turns into a time dependent cosmology. The inner horizon, interestingly, survives this limit and turns into a cosmological horizon. We call this peculiar object a Flat Space Cosmology (FSC). This has previously been studied in literature as the shifted-boost orbifold of Minkowski spacetimes \cite{Cornalba:2002fi, Cornalba:2003kd}. 

We have stated that the dual to 3d flat space would be given by a 2d field theory which enjoys the symmetries of a 2d GCFT. Hence, it is of interest to check whether the proposed dual theory can reproduce the properties of the bulk solutions. Given a cosmological horizon, one can, in analogy with black hole horizons, associate a temperature and a Bekenstein-Hawking entropy    to it. Using GCFT techniques, one can reproduce the entropy by a Cardy-like trick and also the temperature \cite{Bagchi:2012xr}. Interestingly, these FSCs also undergo exotic phase transitions \cite{Bagchi:2013lma}. If one heats up 3D flat space, beyond a critical temperature it goes into a FSC and becomes a time-evolving solution from a static one. 

Given the initial success of the dual field theory, it is important that one pushes this programme and subjects the proposed duality to further tests. In the spirit of AdS$_3$/CFT$_2$, it is thus an important step to check the corrections to the comological horizon entropy from the field theory analysis. This is the principal goal of our present paper.  Along the way, we also derive some additional aspects of the field theory dual to 3d flat spacetimes. 

The outline of the paper is as follows. In Sec 2 and Sec 3, we revisit some relevant details of our proposal of flat space holography stressing on the aspects of the Flat Space Cosmology. It is important to mention that while some of the material here forms a review of earlier works, a lot of this material is new. The development of the ``flat" modular transformation is of particular relevance. In Sec 4, we provide details of the counting of states from the dual GCFT perspective and then apply this to the case of FSC first in Einstein gravity. We show how the result emerges as a limit from the extrapolation of 2d CFT results to the inner horizon of the BTZ. We comment on the limit of vanishing horizon size of the FSC. We reproduce our result by looking at a general thermodynamic analysis and then generalise our saddle-point analysis to Topologically Massive Gravity in flat space. We conclude with a summary of our results and a list of possible future directions.

\section{Holography of 3D Flatspace: Dual Field Theory}

\subsection{Asymptotic Symmetries and Flat limits}
As stated in the introduction, in many ways, the birth of the AdS/CFT correspondence can be attributed to the asymptotic symmetry analysis of Brown and Henneaux  for AdS$_3$ \cite{bh}. The ASA of AdS$_3$ is given by two commuting copies of the Virasoro algebra $\L_n$ and $\bL_n$:
\ben{}
[\L_n, \L_m] &=& (n-m) \L_{n+m} + \frac{c}{12} (n^3-n) \delta_{n+m, 0} \crcr
[\bL_n, \bL_m] &=& (n-m) \bL_{n+m} + \frac{\bar{c}}{12} (n^3-n) \delta_{n+m, 0} \crcr
[\L_n, \bL_m]&=& 0 .
\een
where $c = \bar{c} = \frac{3 \ell}{2 G}$ with $\ell$ as the radius of AdS$_3$. The analogous calculation for 3D flat-spacetimes was done in \cite{Barnich:2006av}. The resulting asymptotic BMS$_3$ symmetry algebra is given by:
\ben{GCA}
[L_n, L_m] &=& (n-m) L_{n+m} + c_L (n^3-n) \delta_{n+m, 0} \crcr
[L_n, M_m] &=& (n-m) M_{n+m} + c_M (n^3-n) \delta_{n+m, 0} \crcr
[M_n, M_m] &=& 0.
\een
Flat spacetime is obtained by taking the radius of AdS$_3$ to infinity. This limit is perceived as a group theoretic contraction on the symmetry structure. This can be explicitly seen by doing the following:
\be{}
L_n = \lim_{\e\to0} (\L_n - \bL_{-n} ), \quad M_n = \lim_{\e\to0} \e (\L_n + \bL_{-n} ) \quad \mbox{where} \, \, \e = \frac{G}{\ell}
\ee
We also see that 
\be{cc-bms}
c_L = \frac{1}{12} (c - \bar{c}) =0 \quad \mbox{and} \quad c_M = \frac{\e}{12} (c + \bar{c}) = \frac{1}{4}.
\ee
This is confirmed by an independent asymptotic symmetry analysis \cite{Barnich:2006av}.

From the point of view of the 2d dual field theory, the limit is a contraction of the time direction $t \to \e t, \, x \to x$ in the conformal structure \cite{Bagchi:2012cy}. This is an ultra-relativistic limit of the parent CFT. Interestingly, field theories with \refb{GCA} as their symmetry algebra had been earlier studied in the context of non-relativistic AdS/CFT \cite{Bagchi:2009my, Bagchi:2009pe}. There the contraction was a non-relativistic one ($t \to t, \, x \to \e x$). The magic of two dimensions means that these two apparently different theories are one and the same. This has been termed the BMS/GCA correspondence. The usefulness of the correspondence is that the techniques developed for non-relativistic conformal field theories can now be applied to understand the dual of flat spacetimes. 

\subsection{Aspects of the Dual Field Theory}

In this section we discuss various aspects of the two dimensional dual field theory including the representations, correlation functions, construct the partition function of the theory and go on to describe aspects of the inherited modular transformation. First we start off with an account of the same set of things in the well known case of two dimensional conformal field theories, so that we can make a direct comparison between the two different classes of field theory.   

\subsubsection{Useful facts about 2d CFTs}
We know that the Virasoro representations are labelled by the conformal weights
\be{}
\L_0 |h, \bar{h} \> = h |h, \bar{h} \>, \quad \bL_0 |h, \bar{h} \> = \bar{h} |h, \bar{h} \>
\ee
Primary states are ones which are annihilated by the action of $\L_n, \bL_n$ for $n>0$.
\be{}
\L_n |h, \bar{h} \>_p = \bL_n |h, \bar{h} \>_p =0
\ee
The Virasoro modules are built on these primary states by acting with creation operators $\L_{-n}, \bL_{-n}$. The representations of the Virasoro algebra on the plane are:
\be{}
\L_n = z^{n+1} \p_z, \quad \bL_n = \z^{n+1} \p_\z
\ee
The map from the plane to the cylinder is
\be{P2C}
z = e^{i \w}, \quad \z = e^{i \bw}
\ee
The generators on the cylinder are given by
\be{}
\L_n = e^{in \w} \p_\w , \quad \bL_n = e^{in \bw} \p_{\bw}
\ee
One constructs the two point function on the plane by demanding that the correlation function be invariant under the symmetries of the theory. The two point function of two primary operators $\Phi_1, \Phi_2$ is given by:
\be{2pt-cft-pl}
G^{(2)}_{\mbox{\tiny{CFT-plane}}}(z_1, \z_1, z_2, \z_2) = C (z_1 - z_2)^{-2 h} (\z_1 - \z_2)^{-2 \h} 
\ee
The transformation law for a primary field $\Phi$ under a co-ordinate transformation $(z, \z) \to (\w, \bw)$ is given by
\be{}
\Phi ' (\w, \bw) = \left( \frac{d\w}{dz} \right)^{-h} \left(\frac{d\bw}{d\z}\right)^{-\h} \Phi (z, \z)
\ee
For the specific case of the mapping from the plane to the cylinder, this is given by 
\be{}
\Phi ' (\w, \bw) = A ( e^{-i\w})^{-h} \left(e^{-i\bw}\right)^{-\h} \Phi (z, \z)
\ee
where $A$ is a phase factor. Using this, one finds the correlation functions on the cylinder from the ones computed on the plane. For example, the two point function on the cylinder for two primary operators is given by
\ben{}
G^{(2)}_{\mbox{\tiny{CFT-cyl}}}(\w_1, \bw_1, \w_2, \bw_2) &=& C \left[\frac{e^{i(\w_1 + \w_2)}}{(e^{i\w_1} - e^{i\w_2})^2}\right]^h \left[\frac{e^{i(\bw_1 + \bw_2)}}{(e^{i\bw_1} - e^{i\bw_2})^2}\right]^\h \nonumber \\\
\Rightarrow G^{(2)}_{\mbox{\tiny{CFT-cyl}}}(\w_1, \bw_1, \w_2, \bw_2)&=& C \{ 2 \sin(\w_1 - \w_2) \} ^{-2h} \{ 2 \sin(\bw_1 - \bw_2) \} ^{-2\h} 
\een
This explicitly depends only on the difference of the co-ordinates on the cylinder.

\subsubsection{Representation and 2pt-functions of 2d GCFTs}

The 2d GCFT representations are labelled by the weights \cite{Bagchi:2009ca}:
\be{l0m0}
L_0 |h_L, h_M \> = h_L |h_L, h_M \>, \quad M_0 |h_L, h_M \> = h_M |h_L, h_M \>
\ee
We will build on the notion of primary states in direct analogy with 2d CFTs. These are ones which are annihilated by the action of $L_n, M_n$ for $n>0$.
\be{}
L_n |h_L, h_M \>_p = M_n |h_L, h_M \>_p =0
\ee
The GCA modules are built on these primary states by acting with creation operators $L_{-n}, M_{-n}$. There is a representation which we would find particularly useful and we will call this the representation on the ``plane"
\be{}
L_n = x^{n+1} \p_x + (n+1) x^n t \p_t , \quad M_n = x^{n+1} \p_t
\ee
The other set of generators, the ones on the cylinder, are our usual ones. These are the generators of field theory which can be read off from the asymptotic analysis on the gravitational side. 
\be{}
L_n = i e^{in \phi} (\p_\phi + in \tau \p_{\tau}), \quad M_n =  i e^{in \phi} \p_\tau
\ee
The map between these two sets of generators or the map between the ``plane" and the ``cylinder" is given by
\be{pl2cyl}
x= e^{i \phi}, \quad t = i \tau e^{i \phi}
\ee
For the plane, we can follow the analysis in \cite{Bagchi:2009ca} to derive the two-point correlation function of the 2d GCFT. This is given by 
\be{gca-2pt-pl}
G^{(2)}_{\mbox{\tiny{GCFT-plane}}}(x_1, t_1, x_2, t_2) = C (x_1 - x_2)^{-2 h_L} \exp{\left[ - 2 h_M \left( \frac{t_1 - t_2}{x_1-x_2} \right) \right]}
\ee
Now we postulate the transformation law for the primary fields for the specific case of the mapping from the ``plane" to the ``cylinder". 
\be{gca-pr-trans}
\Phi ' (\phi, \tau) = A ( e^{-i\phi})^{-h_L} \left(e^{-i\tau}\right)^{-h_M} \Phi (x, t)
\ee
Using this, we can deduce the correlation functions on the ``cylinder".
\be{gca-2pt-cyl-1}
G^{(2)}_{\mbox{\tiny{GCFT-cyl}}}(\phi_1, \tau_1, \phi_2, \tau_2) = C \left[\frac{e^{i(\phi_1 + \phi_2)}}{(e^{i\phi_1} - e^{i\phi_2})^2}\right]^{h_L} \exp \left[ - 2h_M (\tau_1 - \tau_2) \left(\frac{e^{i \phi_1} + e^{i \phi_2}}{e^{i \phi_1} - e^{i \phi_2}}\right)  \right]
\ee
which can be simplified to 
\be{gca-2pt-cyl-2}
G^{(2)}_{\mbox{\tiny{GCFT-cyl}}}(\phi_1, \tau_1, \phi_2, \tau_2) = C_1 \left( 2 \sin(\frac{\phi_{12}}{2}) \right)^{-2h_L} e^{- h_M\tau_{12} \cot (\phi_{12} /2)}
\ee
Interestingly, this is the same answer as one would have got by scaling the 2d CFT 2pt-function in the ultra-relativistic limit \cite{Bagchi:2012cy} (without having to do the extra rescalings to render the answer finite as we needed to do in \cite{Bagchi:2012cy}). 

\subsubsection{Partition Function and Modular Invariance} \label{Z}
In the flat-space limit described above, $(h_L, h_M)$ are mapped to the original eigenvalues of $\L_0, \bL_0$, $(h, \bar{h})$ by
\be{}
h_L = h - \bar{h}, \quad h_M = {\e} (h + \bar{h}).
\ee
In the analysis of the Cardy-like formula, we start with the CFT partition function and re-write it in the ``GCFT-basis". 
\be{pf-cft}
Z_{\textrm{\tiny CFT}} = {\mbox{Tr}} \, e^{2 \pi \zeta \L_0} e^{-2\pi \bar{\zeta} \bL_0} =  \sum d_{{\textrm{\tiny CFT}}} (h, {\bar{h}}) e^{2 \pi i (\zeta h-{\bar{\zeta}} {\bar{h}})}= \sum d(h_{\textrm{\tiny L}}, h_{\textrm{\tiny M}}) e^{2 \pi i (\eta h_{\textrm{\tiny L}}+\frac{\rho}{\e} h_{\textrm{\tiny M}})} 
\ee
$\zeta, \bar\zeta$ are the modular parameters of the original 2d CFT. Above we have relabelled 
\be{etarho}
2\eta = \zeta + \bar \zeta. \quad 2\rho = \zeta - \bar \zeta
\ee
We demand that the partition function of the parent CFT reduce to the GCFT partition function smoothly. 
This implies that $\rho$ has to scale for \refb{pf-cft} to stay finite in the limit. 
\be{}
Z_{\textrm{\tiny CFT}} \xrightarrow{\e\to0} Z_{\textrm{\tiny GCFT}} \Rightarrow \rho \to \e \rho
\ee
We note here that $\rho$ is the variable associated with $M_0$. $M_0$ is the generator of spacetime time translations and hence the Hamiltonian. This is scaled in the limit and it necessitates the scaling of $\rho$ which behaves like the temperature.

Now, modular transformation in original CFT read:
\be{}
\zeta \to \frac{a \zeta + b}{c \zeta + d} \quad \mbox{with} \, \, ad-bc=1
\ee 
In the GCFT basis this translates to:
\be{}
\eta+ \rho \to \frac{a (\eta+ \rho) + b}{c (\eta+ \rho) + d} \to \frac{a \eta + b}{c \eta + d} + \frac{(ad -bc) \rho}{(c \eta + d)^2} + \frac{(ad -bc) c \rho^2}{(c \eta + d)^3} + \ldots
\ee
In the limit, with the scaling of $\rho$, the contracted version of the modular transformation reads
\be{mod-flat}
\eta \to \frac{a \eta + b}{c \eta + d} \qquad \rho \to \frac{\rho}{(c \eta + d)^2} 
\ee
The S-transformation in the original CFT is $\zeta \to - \frac{1}{\zeta}$ and $\bar\zeta \to - \frac{1}{\bar \zeta}$.
This corresponds to $a=d=0$ and $b=-c=1$. So this means that the S-transformation in 2d GCFT reads
\be{S-gca1}
(\eta, \rho) \to \left(- \frac{1}{\eta}, \frac{\rho}{\eta^2}\right)
\ee
This form of the S-transformation has been previously derived in \cite{Bagchi:2012xr}. 
The interesting feature of the full contracted modular transformation is that the second modular parameter $\rho$ is a SL(2,Z) modular form of weight -2. This is reminiscent of the transformations of the variables of a Jacobi form, the features of which we remind the readers of below \footnote{We would like to thank Suresh Govindrajan for this observation and for help with trying to understand the implications.}. 

A Jacobi form $\mathcal{J}$ is a function of two complex variables $\mathcal{J}(\tau, z)$ on which two kinds of transformations act. Under an SL(2,Z) transformation 
\be{tauz}
\tau \to \frac{a \tau + b}{ c\tau +d}, \quad z \to \frac{z}{c \tau + d}                     
\ee
The second transformation is translations by 1 and $\tau$ under which
\be{2trans}
\tau \to \tau \quad \mbox{and} \quad z \to z + \lambda \tau + \mu            
\ee
with $\lambda$ and $\mu$ taking integral values. We can now compare \refb{tauz} with \refb{mod-flat}, choosing $\tau$ to be real and $z$ to be real (or pure imaginary). This suggests the identifications: 
\be{}
\tau \leftrightarrow \eta \quad \mbox{and} \quad Re(z) \leftrightarrow \sqrt{\rho} \quad (\mbox{or} \, Im(z) \leftrightarrow \sqrt{\rho}).
\ee
It is this that suggests that $Z_{\textrm{\tiny GCFT}}$ is like a Jacobi form with real arguments. We are yet to understand the details of this possible relation. A first step would be to try and understand what the analogue of \refb{2trans} is in our case and to see how $Z_{\textrm{\tiny GCFT}}$ transforms under these transformations. We believe it would be of great interest to pursue this line of work to understand the workings of 2d GCFTs from a stronger analytical point of view. Apart from the applications of 2d GCFTs to 3d flat holography, this would be invaluable to diverse areas like non-relativistic conformal theories that were mentioned in the introduction \cite{Bagchi:2009my, Bagchi:2009pe} and also in the case of tensionless string theory where the 2d GCFT finds use as the residual symmetries on the world sheet after fixing the equivalent of the conformal gauge \cite{Bagchi:2013bga}.

\newpage

\section{Holography of 3D Flatspace: Aspects of the Bulk-side} 

As mentioned in the introduction, one of the most surprising features of 3D gravity is the presence of BTZ black holes in AdS$_3$, despite the lack of propagating degrees of freedom. We begin by considering 3-D Einstein-Hilbert gravity with a cosmological constant
 \be{}
 I={1\over16\pi G}\int d^3x\sqrt{-g}\left(R+{2\over\ell^2}\right).
 \ee
The vacuum solution is AdS$_3$ space-time
 \be{}
 ds^2=-\left(1+{r^2\over\ell^2}\right)dt^2+\left(1+{r^2\over\ell^2}\right)^{-1}dr^2+r^2d\phi^2
 \ee
and the excited states are BTZ black holes
\be{BH metric}
 ds^2=-{(r^2-r_+^2)(r^2-r_-^2)\over r^2\ell^2}dt^2+{ r^2\ell^2\over(r^2-r_+^2)(r^2-r_-^2)}dr^2+r^2\left(d\phi-{r_+r_-\over \ell r^2}dt\right)^2.
 \ee
 where
 \be{horizon radia}
 r_\pm=\sqrt{2G\ell(\ell M+J)}\pm\sqrt{2G\ell(\ell M-J)},
 \ee
$M$ and $J$ are related to the mass and angular momentum of the black hole. The Bekenstein-Hawking entropy is given by
\be{}
 S_{BH}={\pi r_+\over 2G}.
 \ee

\subsection{Flat Space Cosmology: Thermodynamics}

Now we want to study the flat space limit of BTZ black hole \eqref{BH metric} by scaling $\ell\to\infty$. It is clear from \eqref{horizon radia} that in the limit, the outer horizon $r_+$ is pushed to infinity 
\be{}
r_+ \to \ell \hat{r}_+ \quad \mbox{where} \quad \hat r_+=\sqrt{8 G M}
\ee 
and hence the radial coordinate $r$ becomes time-like and the time coordinate $t$ becomes spatial. The inner horizon, interestingly, survives the limit
\be{}
r_- \to r_0=\sqrt{{2G\over M}}J.
\ee
and becomes a cosmological horizon.  The resultant geometry is a cosmological solution with a metric
\be{fsc}
ds^2=\hat r_+^2 d t^2-{r^2 dr^2\over \hat r_+^2 (r^2-r_0^2)}+r^2 d\phi^2-2 \hat r_+ r_0 dt d \phi
 \ee
 We shall call these co-ordinates \refb{fsc} ``Schwarzschild" co-ordinates. Defining new coordinate $v, \theta$ as $dv=dt+ {r^2 dr\over \hat r_+^2 (r^2-r_0^2)},\quad d\theta=d\phi+{r_0 dr\over \hat r_+ (r^2-r_0^2)}$ results in the following metric
 \be{}
 ds^2={\hat r_+^2 (r^2-r_0^2)\over r^2}dv^2-2dv dr+r^2\left(d\theta-{\hat r_+ r_0\over
 r^2}dv\right)^2
 \ee
It is clear that surface $r=r_0$ is a null hypersurface and the Killing  vector $\chi=\partial_v+{\hat r_+\over r_0}\p_\theta$
is normal to it. So $r=r_0$ is a Killing horizon and the surface gravity associated to it is
\begin{equation}
\kappa ^2=-{1\over 2}\nabla^\mu\chi^\nu \nabla_\mu \chi_\nu={\hat r_+^4\over r_0^2}
\end{equation}
Hence we can find the Hawking temperature of the FSC 
\be{TH-fsc}
T_H={\kappa\over 2\pi}={\hat r_+^2\over 2 \pi r_0 }
\ee
The entropy of the FSC is found by applying the Bekenstein-Hawking area law to the cosmological horizon :
\be{ent-fsc}
 S_{\mbox{\tiny{FSC}}} = {2\pi r_0\over 4G}={\pi J\over \sqrt{2GM}}
 \ee
It is interesting to observe that the charges associated with the FSC obey a first law of thermodyanmics. 
 \be{}
 dM=-T_H dS_{\mbox{\tiny{FSC}}} + \Omega_{\mbox{\tiny{FSC}}} dJ
 \ee
where $\Omega ={\hat r_+\over r_0}$ is the angular velocity of the horizon. The curious sign in front of the $TdS$ term is an indication of the first law arising as a limit from the BTZ inner horizon thermodynamics, a point which we come back to later. 

In AdS$_3$, there is the well-known Hawking-Page phase transition between hot AdS and the BTZ black hole \cite{Hawking:1982dh}. It is of interest to understand if there exists a flat-space analogue of this \cite{Bagchi:2013lma}. To this end, we study the free energy of these FSC solutions and we approach the problem from the first law of thermodynamics that we have derived above. Using definitions 
\be{}
S = - \frac{\p F}{\p T} \Big|_{\Omega=\textrm{\tiny const.}}, \quad J = \frac{\p F}{\p \Omega} \Big|_{T=\textrm{\tiny const.}}
\ee
we arrive at a different form of the first law of thermodynamics
\be{fl2}
dF = -S dT + J d \Omega \Rightarrow F = U -TS + J \Omega.
\ee
where the first equation has been integrated to get the second. For the FSC, $U = -M$ is the (non-positive) internal energy and hence the free energy takes the form:
\be{}
F_{\mbox{\tiny{FSC}}} = - M_{\mbox{\tiny{FSC}}} = - \frac{\hat{r}_+^2}{8G}.
\ee 
All of this can be obtained from the canonical partition function by continuing to Euclidean signature. If we take Euclidean hot flat space (HFS) 
\be{hfs}
ds_{\mbox{\tiny{HFS}}}^2 = d \tau_E^2 + dr^2 + r^2 d \phi^2.
\ee
at the same temperature and angular potential (periodicities $(\tau_E, \phi) \sim (\tau_E + \beta, \phi + \beta\Omega)$), the Euclidean on-shell action yields the free-energy 
\be{}
F_{\mbox{\tiny{HFS}}} = - \frac{1}{8G} 
\ee
Thus, we can conclude that there is competition between these two solutions in parameter space and a phase transition occurs from HFS to FSC as $\hat{r}$ increases beyond 1 i.e. the temperature increases beyond $T_c = \frac{1}{2 \pi r_0}$ or the angular velocity increases beyond $\Omega_c = \frac{1}{r_0}$. This is a unique phase transition where a time-independent solution (HFS) evolves into a time-dependent cosmological solution \cite{Bagchi:2013lma}.

\subsection{FSC as Minkowski Orbifold}
The family of BTZ black holes correspond to discrete quotients of AdS${}_3$. As we have seen above, the FSC corresponds to a flat limit of the non-extremal BTZ solutions. So these should correspond to discrete quotients of 3d flat space. Let us examine this in a bit more detail. AdS${}_3$ is described by the surface 
\be{emb}
-u^2-v^2 +x^2+y^2 = \ell^2
\ee 
in embedding coordinates in $R^{2,2}$ with metric $ds^2 = -du^2 - dv^2 + dx^2+ dy^2$. Equation \refb{emb} can be solved by setting 
\be{}
  v=\ell\cosh \rho\cos\tau,\,\,\, u = \ell\cosh\rho\sin\tau,\,\,\, x_i = \ell \sinh\rho \, \Omega_i
\ee
which gives the global AdS$_3$ metric $ds^2 = \ell^2 \left( -\cosh^2 \rho \, d\tau^2 + d \rho^2 + \sinh^2 \rho \, d\Omega^2 \right ). $ Non-extremal BTZ black holes are quotients of AdS${}_3$ generated by the Killing vector field
\begin{equation}
  \xi = \frac{r_+}{\ell} J_{ux} - \frac{r_-}{\ell} J_{vy}\, \quad  \mbox{where}  \quad J_{ux} = x\partial_u + u\partial_x,\,\,\,\, J_{vy} = y\partial_v + v\partial_y
\end{equation}
The flat limit corresponds to
\begin{equation}
  \ell\to \infty \,\,\, \tau\to \frac{T}{\ell},\,\,\,\, \rho\to \frac{r}{\ell}
\end{equation}
In this limit, this vector field becomes a linear combination of a boost and a transverse translation
\be{xi}
  \xi_{\text{flat}} = \hat{r}_+ (x \p_T + T \p_x) + r_0\partial_y ,
\ee
where $\hat{r}_+ = \sqrt{2GM}$. One needs to take care of an extra minus sign coming from the fact that the patch that is relevant for the flat limit is the the region $r_- < r < r_+$ region of the original non-extremal BTZ. This coordinate patch has an extra sign when embedding it into $R^{2,2}$.  This is called the shifted-boost orbifold of Minkowski spacetime. 

To understand that this orbifold is indeed the FSC, we can make the following coordinate transformations:
\be{}
x^2 = \frac{r^2 - r_0^2}{\hat{r}_+^2} \sinh^2 (\hat{r}_+ \phi), \quad T^2 = \frac{r^2 - r_0^2}{\hat{r}_+^2} \cosh^2 (\hat{r}_+ \phi), \quad y = r_0 \phi - \hat{r}_+ t 
\ee
It is clear from this that the orbifold \refb{xi} under the change of co-ordinates goes to 
\be{}
\xi_{\text{flat}} = \hat{r}_+ (x \p_T + T \p_x) + r_0\partial_y = \p_\phi
\ee
In the ``Schwarzschild" co-ordinates \refb{fsc}, the orbifold direction was $\phi$. We have shown here that this indeed is the shifted-boost orbifold and also established how we obtain it as a limit of the AdS orbifold.  

\subsection{Modular Transformations in the Bulk}

We have seen the form of the contracted modular transformation in the 2d GCFT \refb{S-gca1}. In the case of AdS$_3$/CFT$_2$, the S-transformations in the dual CFT also have interesting implications for the bulk physics. These transformations map the AdS$_3$ solutions to BTZ black holes. We would like to see if there is a similar mapping of the bulk solutions in 3D flat space \cite{Bagchi:2013lma}. For this we must first continue to Euclidean signature. As stated earlier, the Euclidean version of hot flat space is given by the metric \refb{hfs}. We need also the Euclidean continuation of FSC \refb{fsc}. A natural choice is
\be{}
t = i \tau_{\textrm{\tiny E}}  \qquad \hat r_+ = - i r_+
\ee
which leads to Euclidean FSC metric
\be{Efsc}
d s^2_{\textrm{\tiny E}} = r_+^2\big(1-\tfrac{r_0^2}{r^2}\big) d\tau_{\textrm{\tiny E}} ^2 + \frac{d r^2}{r_+^2(1-\tfrac{r_0^2}{r^2})} + r^2\big(d\varphi - \frac{r_+ r_0}{r^2} d\tau_{\textrm{\tiny E}}\big)^2\,.
\ee
Requiring the absence of conical singularities on the FSC horizon fixes the periodicities of the angular coordinate $\varphi$ and Euclidean time $\tau_{\textrm{\tiny E}}$:
\be{Per}
  \tau_{\textrm{\tiny E}} \sim \tau_{\textrm{\tiny E}} + \frac{2\pi r_0}{r_+^2} = \tau_{\textrm{\tiny E}} + \beta \quad \varphi \sim \varphi + \frac{2\pi}{r_+} = \varphi + \beta \Omega\,,
\ee
The above expressions for Hawking-temperature $T=\beta^{-1}=r_+^2/(2\pi r_0)$ and angular velocity $\Omega=r_+/r_0$ agree with their Minkowski counterparts which we discussed earlier. 

The Euclidean periodicities $\beta, \Omega$ are related to the flat space modular parameters $\eta, \rho$ by:
\be{}
\beta = 2 \pi \rho, \quad \Phi=\beta \Omega = 2 \pi \eta.
\ee
We are now in a position to connect FSC and HFS by the flat $S$-transformation \refb{S-gca1}. The $S$-transformation, in terms of the euclidean periodicities reads
\be{p25}
S: \big(\beta,\, \Phi\big) \to \big(\beta',\, \Phi'\big) = \big(\frac{4 \pi^2 \beta}{\Phi^2},\, -\frac{4 \pi^2}{\Phi}\big)\,.
\ee
Starting with the FSC metric \refb{Efsc}, we change coordinates:
\be{a}
r^2 = r_0^2 + r_+^2 r'^2, \quad \tau_{\textrm{\tiny E}} = \frac{\tau'_{\textrm{\tiny E}}}{r_+} - \frac{\varphi' r_0}{r_+^2}, \quad \varphi = \frac{\varphi'}{r_+}
\ee
This yields flat space:
$$\extd s^2 = \extd\tau_{\textrm{\tiny E}}'^2 + \extd r'^2  + r'^2\,\extd\varphi'^2.$$
In terms of these new coordinates, the periodicities read
\be{b}
(\tau_{\textrm{\tiny E}}',\varphi') \sim (\tau_{\textrm{\tiny E}}' - \beta',\, \varphi' + \Phi') \sim (\tau_{\textrm{\tiny E}}',\, \varphi' + 2 \pi) \quad \mbox{with} \quad \beta' = 2 \pi r_0= \frac{4\pi^2 \beta}{\Phi^2}, \, \Phi' = 2 \pi r_+ = -\frac{4 \pi^2}{\Phi}.
\ee
We recognise these as precisely the values obtained from the $S$-transformation \refb{p25}.
Therefore, we can conclude that FSC with periodicities $(\beta, \,\Phi)$ is equivalent to HFS with flat $S$-dual periodicities $(\beta',\, \Phi')$ \cite{Bagchi:2013lma}. This is the flat space analogue of the statement in the AdS$_3$/CFT$_2$ correspondence that thermal AdS$_3$ with modular parameter $\zeta$ is equivalent to a BTZ black hole with $S$-dual modular parameter $-1/\zeta$ \cite{Kraus:2006wn}.

\newpage

\section{Entropy and Log Corrections}

In this section, we focus on the entropy of the bulk solution that we describe so far, the FSC. We will work in terms of counting of states in the dual field theory. For this we would take recourse to methods adopted in \cite{Carlip:2000nv, Carlip:1998qw} and modify the CFT techniques to suit our needs. We first compute the analogue of the Cardy formula and its leading corrections in the dual 2d GCFT and then focus on obtaining the entropy of the FSC.   

\subsection{State Counting in 2d GCFT}
We have derived the partition function for the 2d GCFT in Sec \refb{Z} and also learnt about the contracted modular invariance. Now, we put the two together to arrive a formula for counting states in the 2d GCFT. The basic result would hinge on the invariance of quantity
\be{pf0}
Z^0_{\textrm{\tiny GCFT}} (\eta, \rho) = {\mbox{Tr}} \,\ e^{2 \pi i \eta (L_0 - \frac{c_L}{2})} e^{2 \pi i \rho (M_0 - \frac{c_M}{2})}= e^{\pi i (\eta {c_L} + \rho c_M)} Z_{\textrm{\tiny GCFT}}(\eta, \rho) 
\ee
under the inherited S-transformation of the GCFT \refb{S-gca1}. Here 
\be{pf}
Z_{\textrm{\tiny GCFT}} (\eta, \rho) = {\mbox{Tr}} \,\ e^{2 \pi i \eta L_0 } e^{2 \pi i \rho M_0} = \sum d(h_L, h_M) e^{2 \pi i \eta h_L} e^{2 \pi i \rho h_M}
\ee
Invariance of $Z^0_{\textrm{\tiny GCFT}}$ under the contracted S-transformation reads
\be{}
Z^0_{\textrm{\tiny GCFT}} (\eta, \rho) = Z^0_{\textrm{\tiny GCFT}} \bigg(- \frac{1}{\eta},\frac{\rho}{\eta^2}\bigg)
\ee
The main observation is now that we can translate this into statements for the partition function.
\be{}
Z_{\textrm{\tiny GCFT}} (\eta, \rho) = e^{2 \pi i \eta \frac{c_L}{2} } e^{2 \pi i \rho \frac{c_M}{2}} e^{- 2 \pi i (- \frac{1}{\eta} )\frac{c_L}{2} } e^{- 2 \pi i (\frac{\rho}{\eta^2}) \frac{c_M}{2}} Z_{\textrm{\tiny GCFT}} \bigg(- \frac{1}{\eta},\frac{\rho}{\eta^2}\bigg)
\ee
By doing an inverse Laplace transformation, one can find the density of states which was defined previously above in \refb{pf}.
\be{den}
d(h_L, h_M) = \int d \eta d \rho \,\ e^{2 \pi i {\tilde{f}}(\eta, \rho)} Z \bigg(- \frac{1}{\eta},\frac{\rho}{\eta^2}\bigg).
\ee
where
\be{}
{\tilde{f}}(\eta, \rho) =  \frac{c_L \eta}{2} +  \frac{c_M \rho}{2} + \frac{ c_L}{2\eta} - \frac{c_M \rho}{2\eta^2} - h_L \eta - h_M \rho.
\ee
In the limit of large charges, the above integration \refb{den} can be performed by the method of steepest descents and the value of the integral is approximated by the value of the integrand when the exponential piece is an extremum. The saddle-point approximation used here is valid when one has an integrand with a rapidly varying phase and a slowly varying prefactor. So, one assumes that the partition function is slowly varying at the extremum. This can be checked. 
In the limit of large charges, the function ${\tilde{f}}(\eta, \rho)$ is approximated by:
\be{}
{f}(\eta, \rho) = \frac{c_L}{2\eta} - \frac{c_M \rho}{2\eta^2} - h_L \eta - h_M \rho.
\ee
The extremum of this is evaluated and the value at the extremum is given by
\be{max}
{f}^{max}(\eta, \rho) = - i \bigg( c_L \sqrt{\frac{h_M}{2c_M}} + h_L \sqrt{\frac{c_M}{2h_M}} \bigg).
\ee
The Cardy-like formula for the GCA in this limit is given by
\be{cardy1}
S^{(0)} = \ln d(h_L, h_M) =  2\pi\bigg( c_L \sqrt{\frac{h_M}{2c_M}} + h_L \sqrt{\frac{c_M}{2h_M}} \bigg).
\ee

\subsection{Logarithmic corrections}
We now want to compute the leading corrections to \refb{cardy1}. In order to evaluate it by saddle point approximation, let's expand $f$ around its saddle given by:
$$ \eta _0 = i \sqrt{\frac{c_M}{2h_M}} ~~~~~ \rho _0 = \frac{h_L \eta_0 ^3}{c_M} + \frac{c_L \eta_0}{2 c_M}.$$
The expansion takes the form (upto quadratic order):
\begin{eqnarray}\label{expan}
f(\eta, \rho) = && f(\eta _0, \rho _0)+\frac{1}{2}[ f_{\eta \eta} (\eta _0, \rho _0)\left(\eta -\eta _0 \right)^2 + 2 f_{\eta  \rho} (\eta _0, \rho _0)\left(\eta -\eta _0 \right)\left(\rho - \rho _0\right) \nonumber \\
&& + f_{\rho \rho} (\eta _0, \rho _0)\left(\rho -\rho _0\right)^2],
\end{eqnarray}
where subscripts denote partial derivative. $f(\eta _0, \rho _0)$ is given by \refb{max}. It is also straightforward to see that
\be{}
 f_{\eta \eta} (\eta _0, \rho _0) = i \sqrt{\frac{2h_M}{c_M}} \left( 3h_L -\frac{C_1 h_M}{C_2} \right), \, \, f_{\eta  \rho} (\eta _0, \rho _0)=2ih_M\sqrt{\frac{2h_M}{C_2}}, \, \, f_{\rho \rho} (\eta _0, \rho _0) =0 
 \ee
Accordingly the term in the square bracket of \eqref{expan} takes the form : $ iz^2-iw^2$, where
$$
z = \left( \frac{2h_M}{c_M} \right)^{1/4} \left( ( \eta - \eta _0) H + \dfrac{2h_M}{H} ( \rho - \rho_0)\right), \, 
w= \left( \frac{2h_M}{c_M} \right)^{1/4} \frac{2h_M}{H} ( \rho - \rho_0)
$$
where $H = \sqrt{3h_L - \frac{c_L h_M}{c_M}}$. On the other hand the measure in \refb{den} changes as:
\be{}
d \eta d \rho \rightarrow dz dw = c_M \left( \frac{2h_M}{c_M} \right)^{3/2}d \eta d \rho .
\ee
\eqref{den} is thus approximately (upto quadratic correction)
\be{}
d(h_L , h_M)=\mathrm{e}^{2 \pi i f(\eta_0, \rho_0)}\, \frac{1}{c_M}\left( \frac{2h_M}{c_M} \right)^{-3/2} K  \quad
\mbox{where }K = \int dz dw \, \mathrm{e}^{2 \pi( -z^2 +w^2)} Z 
\ee
Here we have assumed the constancy of $Z$ near the saddle. $K$ can be evaluated choosing a proper contour passing through the saddle, however that would give a pure number, independent of the charges. Taking the logarithm, we find the logarithmically corrected entropy
\be{log-corrected}
S = 2 \pi \left(c_L \sqrt{\frac{h_M}{2c_M}} + h_L \sqrt{\frac{c_M}{2h_M}}\right) - \frac{3}{2}\log (\frac{2h_M}{c_M^{1/3}}) + \mbox{constant} = S^{(0)} + S^{(1)}
\ee

\subsection{Specializing to FSC in 3D Einstein Gravity}
In the AdS$_3$/CFT$_2$ correspondence, the mass $M$ and angular momentum $J$ of the BTZ black holes are mapped to the conformal weights $(h \bar{h})$ of the dual 2d CFT by the relations
\be{}
h = \frac{1}{2} (\ell M + J) + \frac{c}{24}, \quad \bar{h} =  \frac{1}{2} (\ell M + J) + \frac{\bar{c}}{24}
\ee
where $c= \bar{c} = \frac{3 \ell}{2G}$. This implies that for the FSC, the mass and angular momentum are mapped to
\be{}
h_M = GM + \frac{c_M}{2} = GM + \frac{1}{8}, \quad h_L = J
\ee
In the limit of large charges, we have $h_L = J ,\, h_M \approx GM$ and $c_L=0, c_M= 1/4$. The log corrected horizon entropy takes the form:
\be{ent-1}
S_{\mbox{\tiny{FSC}}}= \frac{\pi |J|}{\sqrt{2GM}} - \frac{3}{2} \log (2GM) + \mbox{ constant}.
\ee
Notice here that the logarithmic correction is independent of angular momentum. We will have more to say about this later. 
The surface gravity for FSC is given by $ \kappa = \frac{\hat{r}^2}{r_0} = \frac{8GM}{r_0}$. Hence the entropy of the FSC can be written in the more familiar form:
\be{fsc-log}
\boxed{S_{\mbox{\tiny{FSC}}}= \frac{2\pi r_0}{4G} - \frac{3}{2} \log (\frac{2 \pi r_0}{4G}) -\frac{3}{2}\log{\kappa}+ \mbox{ constant}}
\ee
\refb{fsc-log} is the main result of this paper. We see that a very familiar factor of $-\frac{3}{2}$ emerges in front of the logarithmic correction {\footnote{We should mention here that we are performing the calculation in the analogue of the ensemble used by Carlip \cite{Carlip:2000nv} where the mass is held fixed and the momentum is summed over. This \cite{Carlip:2000nv} differs from the analysis of Sen in \cite{Sen:2012dw} by a factor of 2. We expect the same to occur if the analysis of  \cite{Sen:2012dw} is adopted for the FSC.}}.In the next subsection, we would see how is linked to the inner horizon of the parent BTZ black hole. 

\subsection{Connections to BTZ Inner Horizon }
There has been recent interest in associating thermodynamics with the inner horizon of black holes \cite{Cvetic:1997uw, Castro:2012av, Detournay:2012ug}. This is due to some intriguing features viz. for asymptotically flat black hole admitting a smooth extremal limit, the product of the inner and outer horizon-areas seems to depend only on the quantized charges and is independent of the mass. The inner horizon also seems to enjoy a first law of thermodynamics of the form
\be{}
- dM = T_- dS_- - \Omega_- dJ + \ldots
\ee
where all the intensive quantities are computed on the inner horizon. Let us concentrate on the non-extremal BTZ black holes. 
For the outer horizon, we have the usual thermodynamic quantities:
\be{}
M_{\mbox{\tiny{BTZ}}} = \frac{r_+^2 + r_-^2}{8G \ell^2}, \, J = \frac{r_+ r_-}{4G}, \, \kappa_+ = \frac{r_+^2 - r_-^2}{r_+ \ell^2}, \, 
\Omega_+ = \frac{r_-}{r_+ \ell}, \, S^{(0)}_+ = \frac{2 \pi r_+}{4G}
\ee
In the above, $\kappa_+$ is the surface gravity on the outer horizon and is related to the temperature $T_+ = \frac{\kappa}{2 \pi}$. These quantities satisfy a first law of thermodynamics:
\be{first-outer}
dM_{\mbox{\tiny{BTZ}}} = T_+ dS^{(0)}_+ - \Omega_+ dJ
\ee
On the inner horizon, we have 
\be{}
\kappa_- = \frac{r_+^2 - r_-^2}{r_- \ell^2}, \, \Omega_- = \frac{r_+}{r_- \ell}, \, S^{(0)}_- = \frac{2 \pi r_-}{4G}
\ee
and this leads to an inner horizon first law:
\be{}
-dM_{\mbox{\tiny{BTZ}}} = T_- dS^{(0)}_- - \Omega_- dJ
\ee

The corrections to the Bekenstein-Hawking area law is found by extending Cardy's analysis in the dual 2d CFT. For the outer horizon, this is the celebrated work of Carlip which we have followed in our GCFT analysis above. Here we quote the result.
\be{S+}
S_+ = S^{(0)}_+ + S^{(1)}_+ = \frac{2 \pi r_+}{4G} - \frac{3}{2} \log \left(\frac{2 \pi r_+}{4G}\right) - \frac{3}{2} \log {\kappa_+} + \mbox{const}
\ee
If we extrapolate this result to the inner horizon, the expected form for the log-corrected entropy for the inner horizon would be
\be{S-}
S_- = S^{(0)}_- + S^{(1)}_- = \frac{2 \pi r_-}{4G} - \frac{3}{2} \log \left(\frac{2 \pi r_-}{4G}\right) - \frac{3}{2} \log {\kappa_-} + \mbox{const}
\ee
where $\kappa_-$ is defined above. 

For the FSC, the various thermodynamic quantities are given by 
\be{}
M_{\mbox{\tiny{FSC}}} = \frac{{\hat r}_+^2}{8G}, \, \Omega_{\mbox{\tiny{FSC}}} = \frac{{\hat r}_+}{r_0}, \, \kappa = \frac{{\hat r}_+^2 }{r_0}, \, S^{(0)}_{\mbox{\tiny{FSC}}} = \frac{2 \pi r_0}{4G}
\ee
These follow a first law of thermodynamics:
\be{}
-dM_{\mbox{\tiny{FSC}}} = T_{\mbox{\tiny{FSC}}} dS^{(0)}_{\mbox{\tiny{FSC}}} - \Omega_{\mbox{\tiny{FSC}}} dJ
\ee
It is easy to see that all of this descends in the limit of $\ell \to \infty$ from the thermodynamic quantities of the inner horizon. It is thus also very plausible that the logarithmic corrections to the FSC entropy would also descend in a similar fashion. It is clear from \refb{S-} and \refb{fsc-log} that indeed this expectation is met. 
\be{}
S_{\mbox{\tiny{FSC}}}=S^{(0)}_{\mbox{\tiny{FSC}}} + S^{(1)}_{\mbox{\tiny{FSC}}}. \qquad
S^{(0)}_- \xrightarrow{\ell\to\infty} S^{(0)}_{\mbox{\tiny{FSC}}}, \, \, S^{(1)}_- \xrightarrow{\ell\to\infty} S^{(1)}_{\mbox{\tiny{FSC}}}
\ee

\subsection{The curious $r_0 \to 0$ limit}

The $r_0 \to 0$ limit of the FSC is the boost orbifold of Minkowski spacetime. This is the limit where the angular momentum is switched off and there is no cosmological horizon to hide the singularity of the causal structure. This can be viewed as a toy Big-Bang model in 3d gravity. As there is no horizon here to hide the singularity, we expect that there would be some funny features in the $r_0 \to 0$ limit. 

First, let us focus on the integrated form of the first law which gives the free energy \refb{fl2}. We observe a peculiar feature of the FSC
\be{}
TS- J\Omega = 0 
\ee
and this holds for all values of $r_0$ and also is true for $r_0 \to 0$. The free energy of the boost orbifold obtained as a $r_0 \to 0$ limit of the FSC thus is the same as the FSC.
\be{}
F_{\mbox{\tiny{boost-orb}}} = -\frac{\hat{r}_+^2}{8G}
\ee
This would mean that in the $r_0 \to 0$ limit, the phase transition between hot flat space (now without a rotation parameter) and the boost orbifold still exists. So now, this is a phase transition between hot non-rotating flat space and a big-bang like cosmological solution.  

There are some obvious peculiarities one observes right away. The temperature previously defined for the FSC \refb{TH-fsc} does not have a well-defined $r_0 \to 0$ limit which is obvious because this was defined in terms of the surface gravity at the horizon which does not exist anymore. Same is true for the angular velocity. This indicates that the phase transition should be characterised not in terms of a critical temperature or a critical angular velocity, but in terms of their ratio which stays finite and non-zero in the limit. In terms of the Euclidean periodicities \refb{Per}, this ratio is precisely the periodicity of the $\phi$ direction. We also observe from \refb{Per} that $\tau_E$ is now trivially identified. So from a thermodynamic perspective, it perhaps makes more sense to think of the angular direction $\phi$ as the new time direction and the inverse of its periodicity $\Phi$ as the temperature. 

We should also point out a possible caveat in the expression of the free energy. We see that the boost orbifold has a singularity in the causal structure which is not hidden by a horizon. In the Euclidean solution one cannot demand smoothness to close off the manifold as one did with the FSC. One can take into account the $r=0$ conical singularity in the value of the free energy. This is something that we would not be able to obtain from a $r_0 \to 0$ limit of the FSC calculations. So it is possible that the limit of vanishing horizon size of the FSC does not capture the entirety of the physics of the boost orbifold. 

Another very curious observation is about the logarithmic corrections to the entropy in the $r_0 \to 0$ limit. It is clear from \refb{ent-1} that the log correction to the FSC entropy is actually independent of $r_0$. So this is going to remain the same in the $r_0 \to 0$ limit. This is very peculiar, since the leading Bekenstein-Hawking entropy in this limit vanishes with the vanishing horizon size. So we are in a domain where the first non-zero piece of the entropy of a gravitational solution is the logarithmic correction. We also have a solution without a horizon, so it is far from clear what this non-zero entropy means. The most plausible solution to this is that our entropy analysis does not hold in the limit that we are presently considering. Indeed, the whole Cardy-like analysis of the 2d GCFT, which is dual to 3D flat space, would be valid in the limit of large charges and we are venturing into a region of parameter space where one of the charges (the angular momentum) is zero. 

Having made some preliminary and speculative remarks about the boost orbifold, we postpone a detailed study of this to another piece of work. The utility of the exercise we have carried out in this sub-section is to convey to the reader that the $r_0 \to 0$ limit is a rather subtle limit to take and there can be many pitfalls along the way.  

\subsection{Entropy and Log corrections from General Thermal Fluctiations}
It was shown in \cite{Das:2001ic} that logarithmic corrections arise in thermodynamic systems when small fluctuations around equilibrium are taken into account. In the microcanonical ensemble, when the leading thermodynamic entropy is given by $S^{(0)}$, the total entropy has a form
\be{thrm}
S = S^{(0)} - \frac{1}{2} \ln (C \, T^2) + \ldots 
\ee
where C is the specific heat of the thermodynamic system. This was successfully applied to the BTZ black hole to reproduce the correct coefficient in front of the logarithmic corrections \cite{Das:2001ic}. We wish to check if this goes through for our FSC solution {\footnote{We would like to thank Daniel Grumiller for suggesting this exercise}}. For the FSC in Einstein gravity, we have \cite{Bagchi:2013lma} 
\be{}
S^{(0)}_{\mbox{\tiny{FSC}}}  = \frac{2\pi r_0}{4G} = \frac{\pi^2}{G} \frac{T_{\mbox{\tiny{FSC}}}}{\Omega^2}, \quad C_{\mbox{\tiny{FSC}}} = T \left( \frac{\p S}{ \p T} \right)\Big|_{\Omega=\textrm{\tiny const.}} = S^{(0)}_{\mbox{\tiny{FSC}}}.
\ee
Putting this into \refb{thrm}, we see that 
\be{}
S_{\mbox{\tiny{FSC}}}=S^{(0)}_{\mbox{\tiny{FSC}}} - \frac{1}{2} \ln \left[ S^{(0)}_{\mbox{\tiny{FSC}}} \frac{G^2 \Omega^4 (S^{(0)}_{\mbox{\tiny{FSC}}})^2}{\pi^4} \right] + \ldots =  S^{(0)}_{\mbox{\tiny{FSC}}} - \frac{3}{2} \ln S^{(0)}_{\mbox{\tiny{FSC}}} + \ldots
\ee
We see that we have been able to reproduce the previous $- \frac{3}{2}$ fraction in front of the logarithmic term. This provides an independent cross-check of the correctness of the answer \refb{fsc-log} obtained by the saddle-point analysis. We can neglect the terms depicted in the ``$\ldots$" when $\Omega \sim 1$, that is when $\hat{r}_+, r_0$ are both of the same order of magnitude. 

\subsection{FSC in Topologically Massive Gravity}
We have so far restricted our attention to Einstein's theory in the 3D bulk. As in the AdS$_3$ example \cite{Li:2008dq}, in flat space the asymptotic symmetries remain the same when one adds a gravitational Chern-Simons term and considers Topologically Massive Gravity (TMG) instead of Einstein gravity \cite{Bagchi:2012yk}: 
\be{}
I_{\tiny{TMG}} = \frac{1}{16\pi G}\, \int d^3 x \sqrt{-g} \left( R + \frac{1}{\mu} CS(\Gamma) \right).
\ee
Here $G$ is the Newton constant, 
$R$ the Ricci scalar, $\mu$ is the Chern--Simons coupling and CS$(\Gamma)=\varepsilon^{\lambda \mu \nu} \Gamma^\rho{}_{\lambda \sigma} \big( \partial_\mu \Gamma^\sigma{}_{\rho\nu} + \frac{2}{3} \Gamma^\sigma{}_{\mu\tau} \Gamma^{\tau}{}_{\nu \rho} \big)$ is the gravitational Chern--Simons term.

The only change occurs in the value of the central term. So, the asymptotic symmetry algebra of TMG in flat space at null infinity is again given by the BMS$_3$ algebra \refb{GCA}, now with two non-zero central terms, viz. 
\be{c-tmg}
c_L = \frac{1}{4 \mu G}, \, c_M = \frac{1}{4G}.
\ee
This can again be motivated by a limit from the AdS-TMG example, where the central terms are $c^\pm = \frac{3 \ell}{2G} (1 \pm \frac{1}{\mu \ell})$. Here we have used $\e = \frac{1}{\ell}$ while scaling the AdS results. 

The equations of motion of TMG amount to the vanishing of the Cotton tensor and this automatically includes all solutions to the Einstein equations. Specifically, the FSC solutions that we have talked about in Einstein gravity also are solutions in TMG. On the gravity side, one can use techniques in \cite{Solodukhin:2005ah, Tachikawa:2006sz} to calculate the entropy of these FSC solutions in TMG. The answer that one gets is \cite{Bagchi:2013lma}
\be{tmg-ent}
S_{\mbox{\tiny{TMG}}}^{\mbox{\tiny{FSC}}(0)} = \frac{2 \pi r_0}{4G} + \frac{2 \pi \hat{r}_+}{4 \mu G}
\ee
The charges $M, J$ in TMG change from their values in Einstein gravity. In the AdS$_3$ case, the changes are \cite{Moussa:2003fc, Olmez:2005by}
\be{}
M \to M + \frac{\mu}{\ell^2} J, \quad J \to J + \frac{1}{\mu} M
\ee
For the flat case, the change in the charges are as follows:
\be{}
M \to M, \quad J \to J + \frac{1}{\mu} M
\ee
So, for the dual theory to TMG in flatspace, states are labelled by the GCFT weights are given by
\be{h-tmg}
h_M = M + \frac{1}{8G}, \quad h_L = J + \frac{1}{\mu} M
\ee
We can now plug the values \refb{c-tmg} and \refb{h-tmg} into the 2d GCFT entropy formula \refb{cardy1} and we can check that in the limit of large charges the entropy formula correctly reproduces \refb{tmg-ent} 
\be{}
S^{(0)}_{\mbox{\tiny{GCFT}}} =  2\pi\bigg( c_L \sqrt{\frac{h_M}{2c_M}} + h_L \sqrt{\frac{c_M}{2h_M}} \bigg) = \frac{2 \pi r_0}{4G} + \frac{2 \pi \hat{r}_+}{4 \mu G} = S_{\mbox{\tiny{TMG}}}^{\mbox{\tiny{FSC}}(0)}
\ee
We can also compute the log corrections to the TMG FSC entropy by looking at \refb{log-corrected}. 
\be{}
S^{{\mbox{\tiny{FSC}}}(1)}_{\mbox{\tiny{TMG}}} = - \frac{3}{2} \log (2GM).
\ee
The intriguing feature is that the log term in the entropy does not depend on the parameter $\mu$ and hence is the same as in Einstein theory. 

\newpage

\section{Conclusions and Future Directions}
In this paper, we have taken further steps to understanding holography in flat spacetimes in three dimensions. Our primary focus in this work was the evaluation of the leading quantum correction to the Bekenstein-Hawking entropy of the Flat Space Cosmological solution of 3D flat space. This logarithmic correction, calculated by looking at the saddle-point analysis to evaluate the density of states in the dual field theory, turned out to be of the form expected from general grounds. By the extrapolation of results from the outer to the inner horizon of the BTZ black hole in AdS$_3$ and taking the flat limit, we showed that the answers obtained matched with those from the dual 2d Galilean Conformal Field Theory. We have also employed techniques for entropy corrections to general thermodynamic systems to cross check our analysis. 

We commented on the curious fact that the result of this analysis did not seem to depend on the radius of the cosmological horizon of the FSC. In particular, in the extreme case where the FSC is non-rotating and hence does not have a horizon, there seems to be a non-zero entropy arising from the solution. Although this is a particularly peculiar and interesting answer, we believe we should not trust it because of the domain of validity of our analysis. 

We have finally looked at extending our analysis to Topologically Massive Gravity in 3D flat space. The putative dual field theory is again a 2D GCFT, now with two non-zero central charges. We showed that the Cardy-like formula for the 2D GCFT reproduces the bulk entropy of the FSC solutions in TMG. We also calculated the log-corrections to the entropy and found that it remained unchanged from its value in Einstein gravity. 

There are numerous unexplored and exciting directions in trying to understand aspects of holography in flat space. Here we list a few which we have directly commented on in this work. First, we would like to devote our attention in the near future to the aspects of the boost orbifold, which is the $r_0 \to 0$ limit of the FSC. One needs to do an independent analysis like in \cite{Bagchi:2013lma} by looking at the Euclidean path integral to compute the on-shell action and the free energy to understand whether the phase transitions to hot flat space exist as we indicated in the $r_0 \to 0$ limit of the final FSC answers in this paper. This is also of importance if one tries to generalise the exotic phase transitions of \cite{Bagchi:2013lma} to higher dimensions. 

Another potentially interesting question is to try and extend the saddle-point analysis beyond the logarithmic correction, following similar the methods to \cite{Loran:2010bd}. One crucial question in this programme is to see if again one can address the $r_0 \to 0$ limit and if there is actually value to the statement that even without a horizon, there exists an entropy which can be associated to the boost orbifold. This would open very interesting questions of the origin of this entropy and whether this can be attributed to the initial time singularity in the causal structure.   

Lastly, we would like to comment on what we believe is an avenue of immense potential. The observation that the modular transformation properties of the partition function of a 2D GCFT are reminiscent of a Jacobi form may help us understand the microscopic details of the dual field theory and thereby provide the basis for an understanding of the counting of states in the field theory at a much more fundamental level. This is an avenue which we would like to make progress on in the near future. 

\bigskip

\section*{Acknowledgements}
Several discussions with Daniel Grumiller, Suresh Govindrajan and Joan Simon are gratefully acknowledged. AB would also like to thank Stephane Detourney, Daniel Grumiller, Reza Fareghbal and Joan Simon for a continuing wonderful collaboration attempting to understand various aspects of holography in flat spacetimes. AB thanks the Institute of Theoretical Physics, Vienna University of Technology and the Chennai Mathematical Institute (CMI) for hospitality during the final stages of this work. A special word of thanks to the organisers and all the participants of the workshop ``Asymptotia" at CMI for all the lively discussions. RB would like to acknowledge the hospitality of IISER, Pune, where the initial part of this work was carried out.  AB is supported by an INSPIRE award by the Department of Science and Technology, India. The research of RB is supported by Department of Science and Technology (DST), Govt. of India research grant under scheme DSTO/1100 (ACAQFT). We would like to thank the people of India for their continued support to research in fundamental sciences.

\bigskip


\end{document}